\documentclass[sigconf]{acmart}

\AtBeginDocument{%
  \providecommand\BibTeX{{%
    \normalfont B\kern-0.5em{\scshape i\kern-0.25em b}\kern-0.8em\TeX}}}

\copyrightyear{2} 
\acmYear{202} 
\setcopyright{rightsretained} 
\acmConference[ACM ]{ACM}{}

\usepackage{pdfpages}
\usepackage{threeparttablex} 
\usepackage{colortbl} 
\usepackage{booktabs}
\usepackage[acronym, toc]{glossaries}
\usepackage[capitalize,noabbrev]{cleveref}
\usepackage{xspace}
\newcommand{\co}{Code Ocean\xspace}
\usepackage{multirow}%
\newcommand{\cross}{\ding{53}\xspace}%
\newcommand{\mycheck}{\ding{51}\xspace}%

\newcommand{\eg}{e.g.\xspace}

\usepackage{verbatim}

\usepackage{mathrsfs}%
\usepackage{pifont}
\usepackage{balance}
\usepackage{hyperref}
\usepackage[capitalize,noabbrev]{cleveref}
\usepackage{url}
\usepackage{tablefootnote}
\usepackage{todonotes}
\usepackage{longtable}
\usepackage{nicematrix}
\usepackage{threeparttable} 

\colorlet{punct}{red!60!black}
\definecolor{background}{HTML}{EEEEEE}
\definecolor{delim}{RGB}{20,105,176}
\colorlet{numb}{magenta!60!black}

\usepackage{tikz}

\def\halfcheckmark{\tikz\draw[scale=0.4,fill=black](0,.35) -- (.25,0) -- (1,.7) -- (.25,.15) -- cycle (0.75,0.2) -- (0.77,0.2)  -- (0.4,0.7) -- cycle;}
\begin{document}

\title{A Dataset For Computational Reproducibility}

\author{Lázaro Costa}
\email{lazaro@fe.up.pt}
\orcid{0000-0002-6317-8792}
\affiliation{
  \institution{University of Porto \&\\ INESC TEC, Portugal}
  \city{}
  \country{}
}

\author{Susana Barbosa}
\orcid{0000-0003-2198-3715}
\email{susana.a.barbosa@inesctec.pt}
\affiliation{%
  \institution{INESC TEC, Portugal}
  \city{}
  \country{}
  }

\author{Jácome Cunha}
\orcid{0000-0002-4713-3834}
\email{jacome@fe.up.pt}
\affiliation{%
  \institution{University of Porto \&\\ HASLab/INESC TEC, Portugal}
  \city{}
  \country{}
  }  

\renewcommand{\shortauthors}{Costa, Barbosa and Cunha}

\begin{abstract}

Ensuring the reproducibility of scientific work is crucial as it allows the consistent verification of scientific claims and facilitates the advancement of knowledge by providing a reliable foundation for future research. 
However, scientific work based on computational artifacts, such as scripts for statistical analysis or software prototypes, faces significant challenges in achieving reproducibility. These challenges are based on the variability of computational environments, rapid software evolution, and inadequate documentation of procedures. 
As a consequence, such artifacts often are not (easily) reproducible, undermining the credibility of scientific findings.

The evaluation of reproducibility approaches, in particular of tools, is challenging in many aspects, one being the need to test them with the correct inputs, in this case computational experiments.

Thus, this article introduces a curated dataset of computational experiments covering a broad spectrum of scientific fields, incorporating details about software dependencies, execution steps, and configurations necessary for accurate reproduction.
The dataset is structured to reflect diverse computational requirements and methodologies, ranging from simple scripts to complex, multi-language workflows, ensuring it presents the wide range of challenges researchers face in reproducing computational studies. It provides a universal benchmark by establishing a standardized dataset for objectively evaluating and comparing the effectiveness of reproducibility 
tools.

Each experiment included in the dataset is carefully documented to ensure ease of use. We added clear instructions following a standard, so each experiment has the same kind of instructions, making it easier for researchers to run each of them with their own reproducibility tool.
The utility of the dataset is demonstrated through extensive evaluations using multiple reproducibility tools. 

\end{abstract}





\maketitle
\section{Introduction} \label{sec:datasetIntroduction}
Reproducibility is a fundamental aspect of the scientific method, essential for validating results and building on existing knowledge~\cite{gundersen2021fundamental, Stodden2010, reproducibility2019report}. However, achieving scientific reproducibility is quite challenging, as reported in numerous studies~\cite{Gavaghan2018, POLDRACK201559}.

Science often relies on some kind of computational work, thus making reproducibility also a computational problem. We term this computational work underlying science \textit{computational experiments}, which we define as collections of files written in any programming language (PL), together with data and other required files or dependencies, necessitating specific configurations within a computational environment to ensure consistent execution.
Despite its importance, achieving reproducibility of computational experiments presents significant challenges. The variability of computational environments (including PLs and their dependencies), rapid software evolution (for instance, new versions of compilers, libraries, or APIs), and inadequate documentation of procedures (from dependencies' management to code execution), leads to difficulties in replicating results. These challenges undermine the credibility of scientific findings and hinders scientific progress~\cite{Nust2021, Brunsdon2021, Ivie2018, DiTommaso2017}.

While multiple tools have been developed to address reproducibility challenges, their effectiveness is difficult to compare due to the absence of common benchmarks~\cite{Peter2022, stodden2014best, Weber2019}. Researchers often struggle to assess these tools consistently, making it challenging to determine their strengths, limitations, and real-world applicability~\cite{costa2024Rep}.

Given these challenges, there is a pressing need for a curated dataset that systematically documents and organizes computational experiments to support reproducibility research~\cite{Stodden2018}. A well-structured dataset must encompass a broad spectrum of scientific fields, ensuring that it reflects the diverse computational requirements and methodologies used across disciplines~\cite{stodden2014best, Weber2019, Peter2022}. Moreover, it should incorporate experiments of varying complexity, ranging from straightforward scripts to intricate, multi-language workflows involving databases and specialized software environments~\cite{Piccolo2016, Stodden2016}.

In addition to covering a wide range of domains and complexities, the dataset should provide explicit details of software dependencies, execution steps, expected results, and any necessary configurations~\cite{Critchlow2013}. This level of documentation is essential for ensuring that experiments can be reliably reproduced in different environments without ambiguity or missing components~\cite{stodden2014best}. Furthermore, the dataset should function as a universal benchmark, establishing a standardized set of experiments for objectively evaluating and comparing reproducibility tools~\cite{Weber2019, Peter2022}. By providing a consistent reference point, researchers can systematically assess tool performance, identify limitations, and drive improvements in reproducibility practices~\cite{Piccolo2016, Stodden2016}.

To address this gap, we propose a curated dataset that reflects the complexity and diversity of modern computational research, covering diverse domains, 
aiming to capture the wide range of challenges researchers face. Each experiment introduces common reproducibility challenges, such as managing software dependencies, handling diverse data formats, and navigating heterogeneous computing environments. By collecting experiments from diverse domains and carefully documenting their characteristics, this dataset provides a set of computational experiments for addressing reproducibility challenges and assessing reproducibility tools' capabilities. We provide more details about the source of the experiments in \cref{sec:datasetCollecting} and fully characterize them in \cref{sec:datasetCaracterization}.

Through our evaluation, we found that 18 out of 38 experiments that were collected could be successfully executed using at least one of the reproducibility tools considered. This 47\% success rate highlights the inherent challenges and limitations in current reproducibility practices, including inconsistent and incomplete documentation for execution steps. 
We present the evaluation and the resulting curated dataset in \cref{sec:datasetEvaluation}.


In \cref{sec:relatedWork}, we discuss related approaches and prior studies that have addressed similar challenges. 
\cref{sec:datasetThreats} analyzes potential limitations in experiment selection, dataset representativeness, and execution conditions. Finally, we discuss our results in \cref{sec:datasetDiscussion} and present the main conclusions in \cref{sec:datasetConclusion}.

\section{Related Work}\label{sec:relatedWork}

Datasets specifically designed for testing and evaluating tools across different fields have played a pivotal role in benchmarking and advancing research. 
For example, the ImageNet Large Scale Visual Recognition Challenge (ILSVRC)~\cite{russakovsky2015imagenet} has been crucial in progressing computer vision by providing a large-scale dataset for benchmarking image recognition algorithms. Similarly, the Penn Treebank~\cite{marcus1993building} has supported advancements in natural language processing by offering annotated corpora for evaluating parsing and tagging tools.
In the realm of image processing and machine learning, the MNIST Database~\cite{deng2012mnist} of handwritten digits has been fundamental. It offers 70,000 labeled images of handwritten digits, widely used for benchmarking image recognition algorithms. Similarly, the COCO (Common Objects in Context) dataset~\cite{lin2014microsoft} has been essential for advancing object recognition technologies by providing complex scene images with detailed object annotations and enhancing tasks like object detection and image captioning.
In the field of natural language processing and social behavior analysis, the Yelp Dataset Challenge~\cite{asghar2016yelp} offers a plethora of user-generated content, business attributes, and social interaction data, enabling a wide range of studies from sentiment analysis to economic forecasting. Additionally, the Microsoft Malware Classification Challenge (BIG 2015)~\cite{ronen2018microsoft} serves the security domain by providing a substantial set of malware samples for developing classification and clustering algorithms.

Although datasets are crucial for many aspects of science, and in particular for evaluations, in the context of reproducibility, there is no such contribution.
Related approaches include earlier efforts, such as SciInc~\cite{Youngdahl2019SciInc} and Sciunit~\cite{Ton2017SciUnits}, focused on proposing reproducibility approaches. 
Although they include a few computational experiments used to evaluate their own contribution, their goal was not to provide a benchmark that could be widely used.
On the other hand, our work significantly extends the evaluation landscape by introducing a comprehensive dataset of 18 experiments designed to challenge and assess reproducibility tools more robustly.

Further advancements were made when researchers evaluated and compared eight key reproducibility tools using a constrained dataset of three experiments \citet{costa2024Rep}. While insightful, this approach was limited by its narrow experimental range, which might not fully capture broader applicability and deeper reproducibility challenges. 
Our paper expands the scope considerably by curating a dataset that spans multiple scientific disciplines, thereby offering a more extensive platform for tool evaluation.


\section{Collecting Experiments} \label{sec:datasetCollecting}

In this section, we define the scientific domains covered, and we describe how we collected the experiments.
In this study, an ``experiment'' is defined as a collection of files written in any PL, encompassing data, configurations, database information, and other necessary files. These components are essential for executing the experiment consistently within a specific computational environment to reproduce the intended results. It is important to note that we exclusively considered experiments that do not depend on specific hardware features such as GPU, or CPU, thus excluding those that require high-performance computing resources. This study included any experiments where it is only necessary to install the required environment, and the results are independent of the hardware used. However, it is important to note that the outcome of these experiments, especially in artificial intelligence (AI) focused on model training, may not be the same in every execution. 

\subsection{Choosing the Experiments}
To develop a robust and diverse dataset, we select a collection of experiments spanning multiple research fields. 

The first is software engineering, represented by experiments published in the \textit{IEEE/ACM International Conference on Software Engineering} (ICSE), a leading venue in the field. The second is databases, with experiments sourced from the \textit{International Conference on Very Large Databases} (VLDB), another highly regarded event in its domain.
In addition, we expanded our scope to include experiments from the field of \textit{AI}; for that, we turned to the Zenodo repository, a well-known platform for open science~\cite{zenodo}.
Moreover, we expanded our scope to include foundational research in software engineering. These studies were sourced from the \textit{European Software Engineering Conference and Symposium on the Foundations of Software Engineering} (ESEC/FSE), a well-regarded venue that frequently features innovative research in software engineering and user-centric methodologies.

Additionally, we collected experiments related to climate change; for that, we explored journals such as \textit{Climate Change}\footnote{\url{https://www.springer.com/journal/10584/}} and \textit{Nature Climate Change}\footnote{\url{https://www.nature.com/nclimate/}}. However, the articles do not provide sufficient details about the computational setup, making it impossible to reproduce the experiments. In this way, we turned to the Zenodo repository to collect experiments in this field.

Furthermore, we considered experiments within the medical and economics domain.
The experiments for both these fields were sourced using Zenodo as our primary platform.

To further enrich the dataset, we explored works addressing reproducibility tools, expecting these would contain some experiments used to illustrate the use of the reproducibility tools. We considered 19 articles reporting reproducibility tools, namely Binder~\cite{Matthias2018binder}, Comprehensive Archiver for Reproducible Execution (CARE)~\cite{Janin2014}, Code, Data, Environment (CDE)~\cite{Philip2011}, Software Provenance in CDE (CDE-SP)~\cite{Pham2015}, \co~\cite{codeocean}, Encapsulator~\cite{Pasquier2018}, FLINC~\cite{Ahmad2022Flinc}, PARROT~\cite{thain2005parrot}, Provenance-To-Use (PTU)~\cite{Pham2013PTU}, Prune~\cite{Ivie2016Prune}, RenkuLab~\cite{ramakrishnan2023renku}, Reprozip~\cite{Chirigati2020}, reprozip-jupyter~\cite{reprojupyter}, ResearchCompendia~\cite{Stodden2015ResearchCompendia}, SciInc~\cite{Youngdahl2019SciInc}, Sciunit~\cite{Ton2017SciUnits}, Science Object Linking and Embedding (SOLE) \cite{malik2014sole}, Umbrella~\cite{Meng2015Umbrella}, and Whole Tale~\cite{Brinckman2019}. 
This ensures the reuse of the largest number of previously used experiments to evaluate existing reproducibility tools.

\subsection{Getting the Experiments}

We began by collecting the experiments published at the three prominent scientific conferences that were chosen. 
Thus, we randomly selected five publications from each of the following conferences: VLDB 2021\footnote{\url{http://vldb.org/pvldb/volumes/15/}}, ICSE 2022\footnote{\url{https://ieeexplore.ieee.org/xpl/conhome/9793835/proceeding?isnumber=9793541}}, and ESEC/FSE 2023\footnote{\url{https://dl.acm.org/doi/proceedings/10.1145/3611643}}. For the last one, we ensure that the selected works have user study experiences.

For the Zenodo repository, we performed four targeted searches using the keywords ``Medical'', ``Artificial Intelligence'', ``Climate Change'', and ``Economics'' for the finding entries related to the respective field. For each search, we filtered results to include only software repositories, focusing on the first 100 entries. From these, we randomly selected five repositories per query to ensure a balanced representation of experiments across the specified domains.

Among the articles reporting reproducibility tools, only two -- \textit{SciInc}~\cite{Youngdahl2019SciInc} and \textit{Sciunit}~\cite{Ton2017SciUnits} -- contained available and complete experiments. From these, we extracted three specific experiments: \textit{Chicago Food Inspections Evaluation}~\cite{e26}, \textit{Variable Infiltration Capacity}~\cite{e27} and \textit{Incremental Query Execution}~\cite{e28}.

\subsection{Summary}

After this process, we were able to collect 38 experiments. Thus, 
we have 13 experiments from software engineering, 
5 from climate change,
5 from medical research,
5 from AI,
5 from economics, and 
5 from user studies. 

\section{Characterization of Collected Experiments} \label{sec:datasetCaracterization}
For each of the 38 experiments collected, we categorized each one in \cref{tab:collectedExperiments}.

We present (in this order) a link to the original repository (clickable in the PDF version of the work), the original publication, the PLs used and the database engine used, the experiment size, the number of files of each experiment, whether the dependencies are described (\mycheck indicates that there is a file with all the details necessary to build the computational environment of the experiment, \cross means that there is not a description, and \halfcheckmark indicates that the dependencies and PLs are listed but without specifying the versions used), and finally the indication of how to execute the experiment (\mycheck indicates that detailed execution steps are provided, \cross indicates insufficient information, ``Package'' is used for repositories that are libraries, packages, or similar artifacts that do not produce a specific result, and ``VS Code Plugin'' is used to mark repositories that are a plugin for VS Code, which are not supposed to produce any particular result and are thus not considered for reproducibility). ``Not available'' is used to mark a repository that is not available, and ``Questionnaire'' is used to mark a repository that only has the content of the questionnaire and no code to execute. Both are not supposed to produce any particular result and are thus not considered for reproducibility therefore, they are excluded from the reproducibility considerations of this study, leaving 36 experiments.

\begin{table*}[htbp]  
\centering
\caption{Characterization of the Collected Experiments}
\label{tab:collectedExperiments}
\rowcolors{2}{white}{gray!25}
\begin{tabular}{m{1.7cm}>{\centering}m{3.3cm}>{\centering}m{4.4cm}>{\centering}m{1.2cm}>{\centering}m{1.05cm}>{\centering}m{1.6cm}>{\centering\arraybackslash}m{2cm}}
\toprule
\rowcolor{gray!50}
    Repo. Link & Pub. & PL/Database & Experiment Size & \# Files & Dependencies Description & How to execute  \\
\midrule

     \href{https://github.com/STAR-Laboratory/Accelerating-RecSys-Training}{\underline{link}} & \citet{Adnan2022} & Unix Shell, Python  & 4.6 GB & 57& \mycheck   & \mycheck \\
     \href{https://github.com/jiyangbai/TaGSim}{\underline{link}} & \citet{Bai2022} & Python & 241.5 MB & 50 & \cross & \mycheck \\
     \href{https://github.com/idea-iitd/RQuBE}{\underline{link}} & \citet{Chauhan2022} & C++
     & 125.6 MB & 71
      & \cross & \mycheck \\
     \href{https://github.com/jt-zhang/CardinalityEstimationTestbed}{\underline{link}} & \citet{Sun2022}  & Python /PostgreSQL  & 364.9 MB & 3636& \cross & \mycheck \\
     \href{https://github.com/AlexanderTZhou/IUBFC}{\underline{link}} & \citet{Zhou2022} & C++& 86.2 MB & 11  & \mycheck & \mycheck \\
     \href{https://github.com/OpsPAI/ADSketch}{\underline{link}} & \citet{Chen2022}  & Python& 2.3 MB & 85     & \mycheck & \mycheck \\
     \href{https://github.com/neu-se/CONFETTI}{\underline{link}} & \citet{Kukucka2022} & Java, Python, Unix Shell& 153.4 MB & 791  & \mycheck & \mycheck \\
     \href{https://github.com/hub-se/BeDivFuzz}{\underline{link}} & \citet{Nguyen2022} & Python, Java  & 2.9 MB & 758& \mycheck & \mycheck \\
     \href{https://github.com/ICSE2022FL/ICSE2022FLCode}{\underline{link}} & \citet{Xie2022}  & Python & 10.4 MB & 66   & \mycheck & \mycheck  \\
     Not available & \citet{Zhang2022} & \multicolumn{5}{c}{Not available} \\
    \href{https://zenodo.org/record/5967578#.ZEwQKnbMJD8}{\underline{link}} & \citet{edwards2022}  & R & 488.7 KB & 69 & \mycheck & Package \\
    \href{https://zenodo.org/record/7651416#.ZEwPxXbMJD8}{\underline{link}} & \citet{Hemes2023}  & R  & 71.4 KB & 3& \mycheck & \cross  \\
    \href{https://zenodo.org/record/4588383#.ZEwRHnbMJD8}{\underline{link}} & \citet{Jehn2021} & Python  & 2.7 MB & 64& \cross & \cross \\
    
    \href{https://zenodo.org/record/7613549#.ZEwQSnbMJD8}{\underline{link}} & \citet{Lin2024} & Python & 128.8 MB & 208 & \cross & \mycheck \\
    \href{https://zenodo.org/record/5233947#.ZFPecHbMJD8}{\underline{link}} & \citet{Qasmi2021} & R & 70.3 KB & 31 & \mycheck & Package \\
     \href{https://zenodo.org/record/4497214#.ZFQy33bMJD-}{\underline{link}} & \citet{Frie2021} & Python & 21.3 MB & 217 & \cross & \mycheck \\
     \href{https://zenodo.org/record/4926118#.ZEFWgnbMJD8}{\underline{link}} & \citet{Tianyu2021}  & Python  & 139.1 KB & 33& \mycheck & \mycheck \\
     \href{https://zenodo.org/record/1488650#.ZFQw4nbMJD_}{\underline{link}} & \citet{Hoyt2018} & Python & 261.8 KB & 67 & \mycheck & Package \\
     \href{https://zenodo.org/record/6617957#.ZEFTEXbMJD8}{\underline{link}} & \citet{Kailas2022}  & Python, Unix Shell & 1 MB & 167  & \halfcheckmark & \mycheck  \\
     \href{https://zenodo.org/record/4453803#.ZFQxW3bMJD-}{\underline{link}} & \citet{Yang2021} & Python & 38.7 MB & 37  & \mycheck & \mycheck \\
     \href{https://zenodo.org/record/7867100}{\underline{link}} & \citet{Prezja2013} & Python & 3.8 MB & 91 & \mycheck & Package \\
     \href{https://zenodo.org/record/7595510}{\underline{link}} & \citet{Santucci2023} & Python & 352.8 KB & 6  & \mycheck & \cross \\
     \href{https://zenodo.org/record/3568468}{\underline{link}} & \citet{Erickson2019} & Python, Unix Shell, C++ & 5.5 MB & 459 & \mycheck & Package\\
     \href{https://zenodo.org/record/6468193}{\underline{link}} & \citet{Coupette2022} & Python, Jupyter Notebook & 114.4 MB & 93 & \halfcheckmark & \cross \\
     \href{https://zenodo.org/record/7585845}{\underline{link}} & \citet{doe2023axom} & Python & 1.9 MB & 25  & \mycheck & \cross \\
    \href{https://chicago.github.io/food-inspections-evaluation/}{\underline{link}} & \citet{e26}  & R & 147.9 MB & 101 & \mycheck & \mycheck \\
    \href{https://github.com/uva-hydroinformatics/VIC_Pre-Processing_Rules}{\underline{link}} & \citet{e27}    & Python, Unix Shell & 254 KB & 60 & \cross & \cross  \\
    \href{https://bitbucket.org/TonHai/iqe/src/master/}{\underline{link}} & \citet{e28}   & Python /SQLite & 68.6 KB & 26& \cross & \cross \\
     \href{https://zenodo.org/records/3980817}{\underline{link}} & \citet{e29}   & Jupyter Notebook & 1.1 GB & 59988     & \mycheck  & \mycheck    \\
     \href{https://zenodo.org/records/4562932}{\underline{link}} & \citet{e30}   & R &   4.6 MB& 31    & \cross  & \cross    \\
     \href{https://zenodo.org/records/4391483}{\underline{link}} & \citet{e31}  & R &  182.9 KB & 15    & \mycheck  & \mycheck    \\
     \href{https://zenodo.org/records/3377463}{\underline{link}} & \citet{e32}  & R, Python, Unix Shell &   327.9 KB& 15   & \mycheck  & \cross    \\
     \href{https://zenodo.org/records/5656923}{\underline{link}} & \citet{e33}  & R & 183.7 MB &   4062  & \mycheck  & \mycheck  \\
     \href{https://github.com/nanofuzz/nanofuzz}{\underline{link}} & \citet{e34}   & TypeScript, JavaScript & 1.1 MB & 81    & \mycheck & VS Code Plugin    \\
     \href{https://zenodo.org/records/8271643}{\underline{link}} & \citet{e35}   & TypeScript, JavaScript, Unix Shell &  98.7 MB& 14733   & \mycheck & VS Code Plugin    \\
    \href{http://r.jyu.fi/writing-task-about-flow}{\underline{link}} & \citet{e36}   & \multicolumn{5}{c}{Questionnaire}  \\
    \href{https://zenodo.org/records/8271853}{\underline{link}} & \citet{e37}   & Python, Unix Shell  & 4 MB & 70    & \halfcheckmark  & \mycheck   \\
     \href{https://zenodo.org/records/8270217}{\underline{link}} & \citet{e38}   & Python &   134.4 MB  &   6409 & \mycheck & \mycheck \\
        \bottomrule
\end{tabular}
\end{table*}

The collected experiments span a wide range of PLs, reflecting the diversity of computational approaches utilized in scientific research. Notably, Python is the most frequently used PL, present in 67\% (24 of 36) of the experiments, underscoring its dominance in scientific computing and data analysis. Unix Shell and R are the second most common, used in 22\% (8 of 36) of the experiments, highlighting their importance in automating workflows and statistical analysis. C++ is utilized in 8\% (3 of 36) of the experiments. Finally, TypeScript, JavaScript, Java, and Jupyter Notebook each represent 6\% (2 of 36).

The size of the collected experiments varies significantly, showcasing its heterogeneity. The smallest experiment has 68.6 KB, while the largest exceeds 4.6 GB, illustrating a broad spectrum of computational complexities. Similarly, the number of files per experiment ranges from as few as three files to as many as 59,988 files, further demonstrating the variation in experiment structure and scope.

In terms of reproducibility, these experiments reveal notable gaps. The data, detailed in \cref{tab:collectedExperiments}, shows that while 67\% (24 of 36) of the experiments include all the details necessary to build the computational environment of the experiment, a significant proportion still lacks this critical information. Furthermore, only 50\% (18 of 36) of the experiments provide detailed guidance on how to execute them, which is essential for ensuring replicability in different computational environments. These gaps highlight the challenges researchers face when attempting to reproduce computational experiments from existing publications~\cite{peng2011reproducible}.

Overall, the collected experiments present a diverse and heterogeneous collection of experiments in terms of PL, size, and file count. However, it also exposes significant shortcomings in the documentation of dependencies and execution instructions, which are vital for advancing reproducibility in scientific research. 

In the following section, we introduce the curated dataset, which showcases experiments designed to function as a universal benchmark.

\section{Evaluating the Collected Experiments} \label{sec:datasetEvaluation}
This section presents the results of our efforts to reproduce the collected experiments described in \cref{tab:collectedExperiments} using eight reproducibility tools: Binder~\cite{Matthias2018binder}, \co~\cite{codeocean}, FLINC~\cite{Ahmad2022Flinc}, PTU~\cite{Pham2013PTU}, RenkuLab~\cite{ramakrishnan2023renku}, Reprozip~\cite{Chirigati2020}, Sciunit~\cite{Ton2017SciUnits}, and Whole Tale~\cite{Brinckman2019}, which were previously identified in a comprehensive study~\cite{costa2024Rep}.

\cref{tab:tool-access} details each tool's location (either a GitHub repository or a dedicated webpage) and whether the tool was accessed online or installed locally. All tools were accessed in October 2024. In the initial phase of executing the experiments, we were able to use the Whole Tale platform. However, when we attempted to add more collected experiments, the platform became unavailable.

Experiments executed locally were run on a personal computer equipped with a 6-core CPU and 16 GB of RAM, operating under Ubuntu 20.04.

\begin{table}[h]
\centering
\caption{Summary of the Tools Used for Evaluating the Dataset}
\label{tab:tool-access}
\rowcolors{2}{white}{gray!25}
\begin{tabular}{m{1.1cm}>{\centering}
m{4.5cm}>
{\centering\arraybackslash}m{1.7cm}}
\toprule
\rowcolor{gray!50} 
\textbf{Tool}       & \textbf{Location}            & \textbf{Access Mode}  \\ \midrule
Binder              & \url{https://mybinder.org/}  & Online\\
\co          & \url{https://codeocean.com/} & Online\\
FLINC               & \url{https://github.com/depaul-dice/Flinc}   & Installed Locally     \\
PTU & \url{https://github.com/depaul-dice/provenance-to-use}       & Installed Locally     \\
RenkuLab            & \url{https://renkulab.io/}   & Online, Version 2.0   \\
Reprozip            & \url{https://www.reprozip.org/}              & Installed Locally     \\
Sciunit             & \url{https://github.com/scidash/sciunit}     & Installed Locally     \\
Whole Tale          & \url{https://wholetale.org/} & Online\\ \bottomrule
\end{tabular}
\end{table}

\subsection{Running the Experiments}
For each experiment that we collected, we installed the necessary dependencies and attempted to reproduce the results using each tool. 
We standardized the preparation and documentation process for each experiment, and we added the file \textit{READ-FOR-REPRODUCIBILITY.md} to each experiment. Although we strived to follow a consistent structure for describing the execution environment, the execution process itself, and how the information is presented, there is currently no universally accepted template for this type of documentation. Therefore, we selected and organized the information we considered most relevant for reproducibility\cite{Leipzig2021}.

The outcomes are summarized in \cref{tab:reprodution}, which includes the reproducibility status and the generated package size (in MB, using the Zip format) when applicable. For failed executions, specific reasons are annotated, such as unsupported PL, compilers, databases, or tool limitations.
For each experiment, we created a reproducibility package with each tool that supported it. These packages include the source code, data, and execution scripts, and are publicly available for review and reuse~\cite{costa_2025_packahe_experiments_anonymous}.

The reproducibility status is labeled as follows: ``\mycheck'' -- the experiment was successfully reproduced, and ``\cross{}'' -- the experiment could not be reproduced. 
We use specific labels to classify repository availability and reproducibility. 
``No Data'' signifies that the necessary data for reproduction is missing from the repository. ``Missing Files'' is used when specific files are absent, leading to execution errors. ``Partial'' applies to repositories that are only partially reproducible, usually because some required data is restricted by licensing. For experiments where the platform became unavailable, we marked those experiments with a ``-''.

\begin{table*}[htbp]  
\caption{Reproducibility of each experiment on each platform and the package size in MB (in parentheses), when applicable}
\rowcolors{2}{white}{gray!25}
\begin{tabular}{m{3.4cm}>{\centering}m{1cm}>{\centering}m{1.5cm}>{\centering}m{1.1cm}>{\centering}m{1.8cm}>{\centering}m{1.3cm}>{\centering}m{1.7cm}>{\centering}m{1.2cm}>{\centering\arraybackslash}m{1.1cm}}
\toprule
\rowcolor{gray!50} 
      Publication & Binder & \co & FLINC & PTU & RenkuLab & ReproZip & Sciunit & Whole Tale  \\
\midrule
        \citet{Adnan2022} & \cross{}\textsuperscript{b} & \mycheck (3900)  & \cross{}\textsuperscript{b} & \mycheck (986.2) & \mycheck & \mycheck (974.4) & \mycheck(269.5) & \cross{}\textsuperscript{b}  \\ 
        \citet{Bai2022} & \cross{}\textsuperscript{a} & \mycheck (241.4)& \cross{}\textsuperscript{b} & \mycheck (2100) & \mycheck & \mycheck (297)& \mycheck (2000) & \cross{}\textsuperscript{b}  \\ 
        \citet{Chauhan2022} & \cross{}\textsuperscript{b} & \cross{}\textsuperscript{c} & \cross{}\textsuperscript{b} & \mycheck (119.6) & \cross{}\textsuperscript{b} & \mycheck (26.4) & \mycheck (61.5) & \cross{}\textsuperscript{b}  \\ 
        \citet{Sun2022} & \cross{}\textsuperscript{d} & \cross{}\textsuperscript{d} & \cross{}\textsuperscript{b} & \mycheck (71.6)& \cross{}\textsuperscript{d} & \mycheck (69.1) & \mycheck (69.2)& \cross{}\textsuperscript{b}  \\ 
        \citet{Zhou2022} & \cross{}\textsuperscript{b} & \mycheck (86.2) & \cross{}\textsuperscript{b} & \mycheck (31.0)& \cross{}\textsuperscript{b} & \mycheck (30.6) & \mycheck (6.9)& \cross{}\textsuperscript{b}  \\ 
        
        \citet{Chen2022} & \mycheck & \mycheck (2.3)& \cross{}\textsuperscript{b} & \mycheck (152.4)& \cross{}\textsuperscript{e} & \mycheck (150) & \mycheck (148.5)& \cross{}\textsuperscript{b}  \\ 
        \citet{Kukucka2022} & \cross{}\textsuperscript{b} & \cross{}\textsuperscript{a} & \cross{}\textsuperscript{b} & \mycheck (1.1)& \cross{}\textsuperscript{b} & \mycheck (225.6) & \mycheck (1.1) & \cross{}\textsuperscript{b}  \\ 
        \citet{Nguyen2022} & \cross{}\textsuperscript{b} & \cross{}\textsuperscript{a} & \cross{}\textsuperscript{b} & \cross{}\textsuperscript{a} & \cross{}\textsuperscript{b} & \cross{}\textsuperscript{a} & \cross{}\textsuperscript{a} & \cross{}\textsuperscript{b}  \\ 
        \citet{Xie2022} & \mycheck  & \mycheck (10.3)& \cross{}\textsuperscript{b} & \mycheck (2100) & \mycheck & \mycheck (2100)& \mycheck (1900) & \cross{}\textsuperscript{b} \\ 
        
        \citet{Hemes2023} & \multicolumn{8}{c}{No Data} \\ 
        \citet{Jehn2021} & \mycheck & \mycheck (2.7)& \cross{}\textsuperscript{b} & \mycheck (75.2)& \mycheck & \mycheck (73.7) & \mycheck  (74.9)& \cross{}\textsuperscript{b}  \\ 
        \citet{Lin2024} & \mycheck & \mycheck (128.8)& \mycheck  (21.2)& \mycheck (685.4) & \mycheck & \mycheck (650.1) & \mycheck (88.9) & \cross{}\textsuperscript{b}  \\ 
        
        \citet{Frie2021} & \multicolumn{8}{c}{No Data} \\ 
        \citet{Tianyu2021} & \multicolumn{8}{c}{No Data} \\ 
        \citet{Kailas2022} & \cross{}\textsuperscript{e} & \cross{}\textsuperscript{e} & \cross{}\textsuperscript{b} & \cross{}\textsuperscript{a} & \cross{}\textsuperscript{e} & \cross{}\textsuperscript{a} & \cross{}\textsuperscript{a} & \cross{}\textsuperscript{b}  \\
        
        \citet{Yang2021} & \cross{}\textsuperscript{a} & \cross{}\textsuperscript{a} & \cross{}\textsuperscript{b} & \cross{}\textsuperscript{a} & \cross{}\textsuperscript{a} & \cross{}\textsuperscript{a} & \cross{}\textsuperscript{a} & \cross{}\textsuperscript{b}  \\ 
        
        \citet{Santucci2023} & \mycheck & \mycheck (0.358) & \cross{}\textsuperscript{b} & \mycheck (77.1)& \mycheck & \mycheck (83.4)& \mycheck (86.6) & \cross{}\textsuperscript{b}  \\ 
        \citet{Coupette2022} &  \multicolumn{8}{c}{No Data} \\
        
        \citet{doe2023axom} & \cross{}\textsuperscript{a} & \cross{}\textsuperscript{a} & \cross{}\textsuperscript{b} & \cross{}\textsuperscript{a} & \cross{}\textsuperscript{a} & \cross{}\textsuperscript{a} & \cross{}\textsuperscript{a}& \cross{}\textsuperscript{b}  \\
        
        \citet{e26} & \cross{}\textsuperscript{a} & \mycheck (147.9) & \cross{}\textsuperscript{b} & \mycheck (33.1) & \mycheck & \mycheck (39.4) & \mycheck (177.7) & \mycheck  \\ 
        \citet{e27} & \multicolumn{8}{c}{Miss Files } \\
        \citet{e28} & \mycheck & \cross{}\textsuperscript{e} & \cross{}\textsuperscript{b} & \mycheck (6) & \cross{}\textsuperscript{e} & \mycheck (5.7) & \mycheck (5.7) & \cross{}\textsuperscript{b}  \\ 
        
        \citet{e29} & \mycheck & \cross{}\textsuperscript{a} & \mycheck (21.2) & \cross{}\textsuperscript{b} & \mycheck & \cross{}\textsuperscript{b} & \cross{}\textsuperscript{b} & -  \\ 
        \citet{e30} & \cross{}\textsuperscript{a} & Partial (4.6)& \cross{}\textsuperscript{b} & Partial (65.5)& Partial & Partial  (67.5)& Partial (71.3) & -  \\ 
        \citet{e31} & \cross{}\textsuperscript{a} & \cross{}\textsuperscript{a} & \cross{}\textsuperscript{b} & \cross{}\textsuperscript{a} & \cross{}\textsuperscript{a} & \cross{}\textsuperscript{a} & \cross{}\textsuperscript{a} & -  \\ 
        
        \citet{e32} & Partial & \cross{}\textsuperscript{e} & \cross{}\textsuperscript{b} & Partial  (0.289)& \cross{}\textsuperscript{e} & Partial (0.184) & Partial (0.284) & -  \\ 
        
        \citet{e33} & \cross{}\textsuperscript{a} & \cross{}\textsuperscript{b}  & \cross{}\textsuperscript{b} & Partial (97.4) & Partial & Partial (99.6) & \cross{}\textsuperscript{a} & - \\ 

        \citet{e37} & \cross{}\textsuperscript{a} & \cross{}\textsuperscript{a} & \cross{}\textsuperscript{b} & \cross{}\textsuperscript{a} & \cross{}\textsuperscript{a} & \cross{}\textsuperscript{a} & \cross{}\textsuperscript{a} & -  \\ 
        \citet{e38} & \mycheck & \mycheck (135.5)& \cross{}\textsuperscript{b} & \mycheck (99.7) & \mycheck & \mycheck (10.8) & \mycheck (99.1) & - \\
        \bottomrule
\end{tabular}
\begin{flushleft}
\textsuperscript{a} Code crashed\\
\textsuperscript{b} PL not supported\\
\textsuperscript{c} g++ compiler not supported\\
\textsuperscript{d} Database not supported\\
\textsuperscript{e} Python version not supported\\
\textsuperscript{f} Tool/IDE not supported
\end{flushleft}
\label{tab:reprodution}%
\end{table*}


We could not run the experiment from \citet{Nguyen2022} in any framework, even in the ones that fully support the Java environment (\co, PTU, Reprozip, and sciunit), because we could not find a Java class required by the experiment. 
For the experiments conducted by \citet{Kailas2022} and \citet{e37}, the required version of Python was not documented. We attempted execution using Python versions 3.5, 3.7, 3.8, and 3.9; however, all were found to be incompatible with several necessary packages, thus preventing successful experiment execution. Additionally, for the experiments conducted by \citet{e31} and \citet{Yang2021}, we encountered unresolved conflicts during the package installation process.
For the experiments conducted by \citet{doe2023axom}, we encountered an unexpected input data format, which rendered execution impossible.



Notably, we were not able to run these experiments on our personal computers either. The problem may be related to a possible error in the source code or some failure in the configuration of the computational environment due to a lack of information. 
Hence, it is crucial to provide not only the source code of the experiment but also all required information for effective execution, including data, details about the computational environment used, and any other relevant execution details. 

For the experiments by \citet{e30}, \citet{e32}, and \citet{e33}, we had access to only a portion of the complete set of data of the experiment. Although we contacted the authors to request full access, we were unable to obtain the necessary license. As a result, we could only partially reproduce the experiment.

From the initial selection of 38 experiments, we encountered several challenges that influenced the final curated dataset to be used to test computational reproducibility tools. 
Five of the experiments in the initial collection were identified as packages or libraries, which cannot be included in a reproducibility study as they do not execute or produce any specific results tied to the corresponding publications. Additionally, six experiments lacked access to either the source code or the required data, rendering them impossible to execute. Furthermore, two experiments were VS Code plugins, which do not produce expected outputs and are not supported by any of the evaluated platforms.
One experiment was a questionnaire containing only the content of the questionnaire itself, which was thus also excluded. 
Finally, six experiments could not be successfully executed on any platform despite meeting other criteria and were therefore excluded from the curated dataset.

\subsection{Curated Datasset}\label{sec:datasetCurated}
In \cref{tab:curatedDataset}, we define the curated dataset of experiments that only includes those that have been successfully reproduced by at least one of the eight reproducibility tools considered. 
This curated selection focuses exclusively on experiments that produce a tangible result, excluding cases such as R packages and plugins, which are not intended to generate specific outcomes. By concentrating on experiments with clear results, this dataset offers a robust benchmark for comparing the efficiency and effectiveness of reproducibility tools.

Additionally, we resolved all issues related to incomplete dependency specifications, insufficient PL description, and incomplete execution steps across the curated dataset.

Thus, the curated dataset consists of 18 original experiment repositories that were successfully reproduced using at least one of the evaluated reproducibility tools. This dataset is available in \cite{costa_2025_dataset_anonymous}. Additionally, this dataset serves as a reference for researchers to explore reproducibility challenges and develop improved methodologies.

As detailed in \cref{sec:datasetCaracterization}, the curated dataset retains the diversity of PL and domains while excluding experiments that failed reproducibility assessments.
It includes experiments implemented in Python, R, C++, and other languages, covering a wide range of domains.
Details of the curated dataset are presented in \cref{tab:curatedDataset}.
Each experiment of the curated dataset is accompanied by a clear link to its repository, the associated publication, and metadata such as the PL and database engines used and the classification following the \textit{Artifact Review and Badging (v1.1)}\footnote{ACM. Artifact Review and Badging (v1.1), accessed on March 30, 2025, \url{https://www.acm.org/publications/policies/artifact-review-and-badging-current}} guidelines from ACM, applied as follows:

\begin{description}
    \item[\textit{Artifacts Evaluated - Functional}]\includegraphics[width=0.4cm]{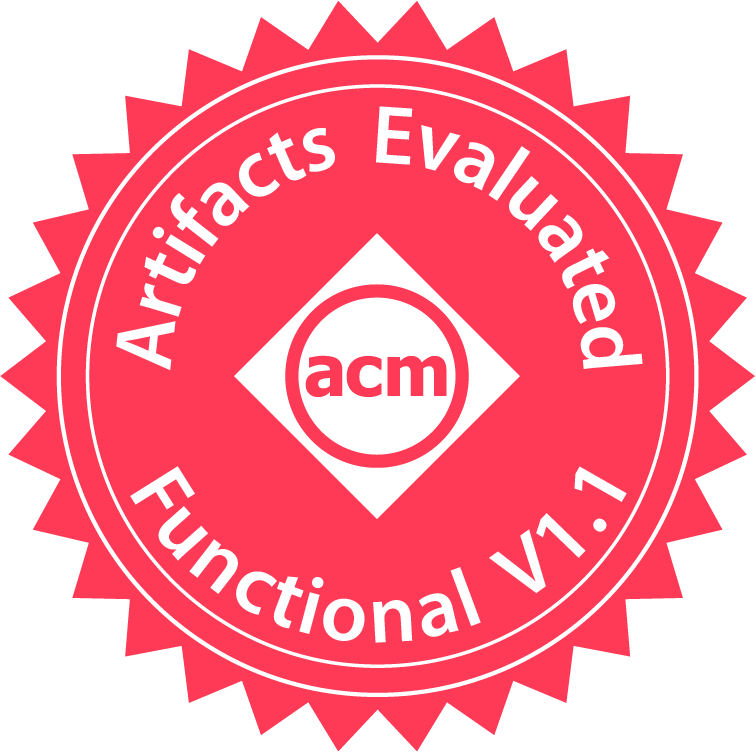} – The artifact has been reviewed and verified to be documented, complete, and functional.
    
   \item[\textit{Artifacts Evaluated - Reusable}] \includegraphics[width=0.4cm]{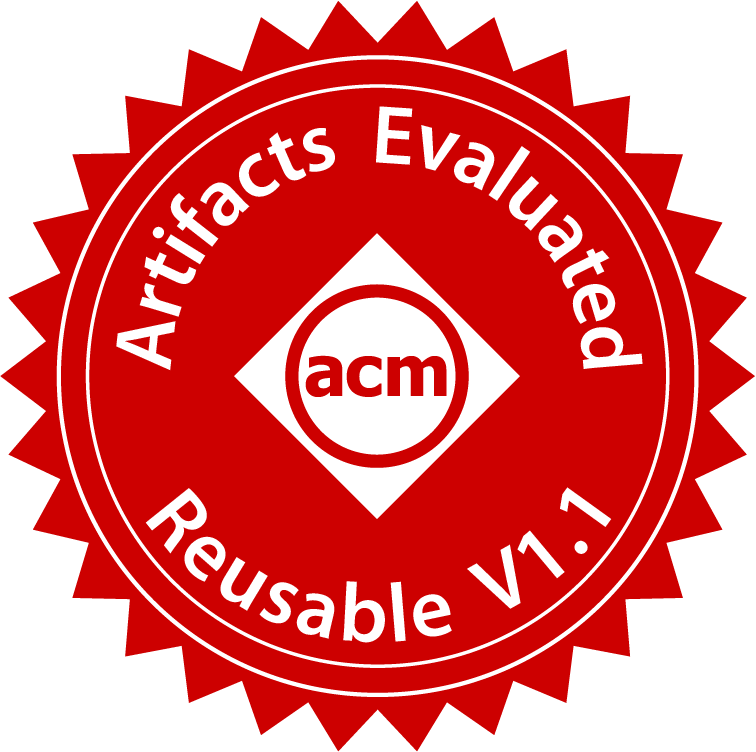}– The artifact provides significant functionality and is well-documented for reuse by other researchers.
    
    \item[\textit{Artifacts Available}]\includegraphics[width=0.4cm]{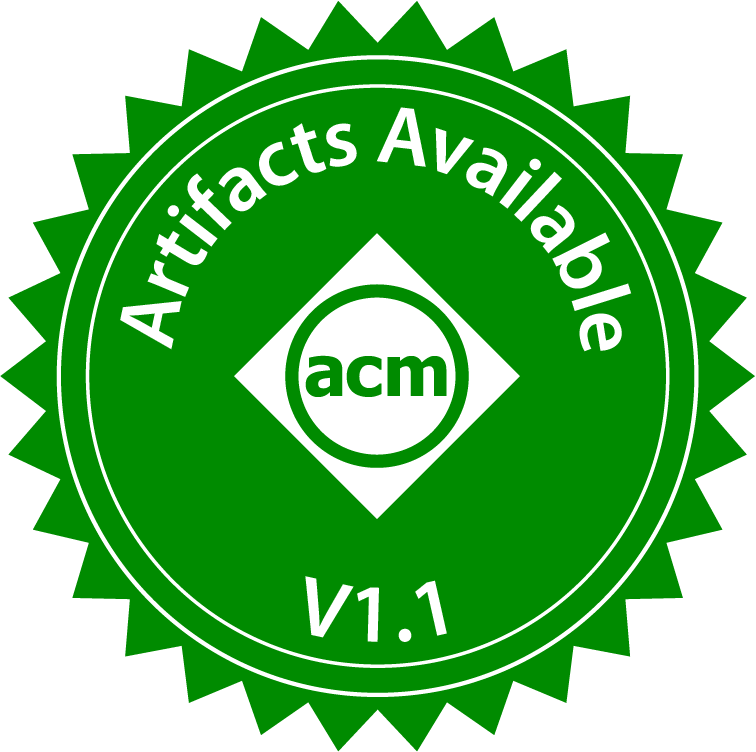} – The artifact is stored in a publicly accessible repository with a unique identifier, ensuring long-term availability.
    
   \item[\textit{Results Validated - Reproduced}]\includegraphics[width=0.4cm]{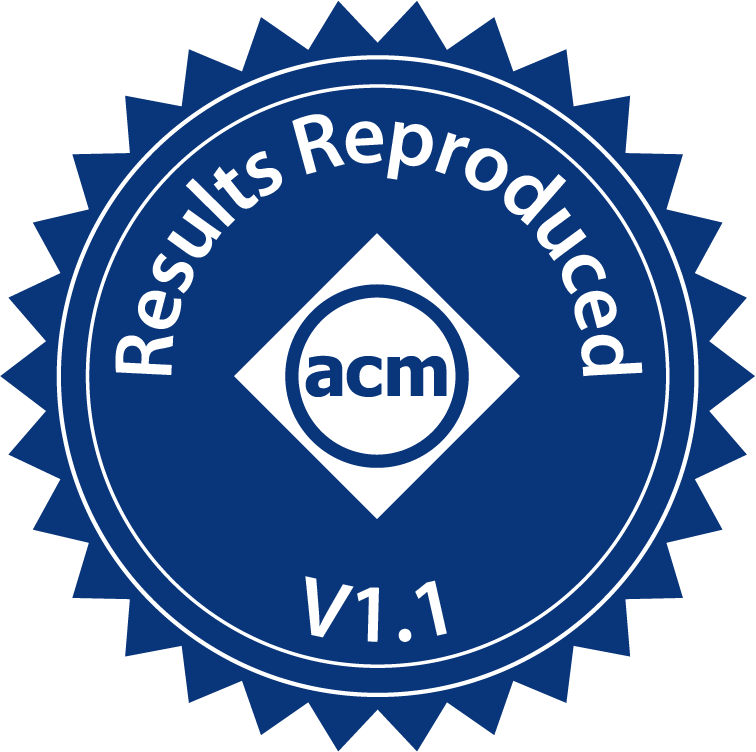} –  This artifact indicates that the main results of the original study have been successfully obtained by an independent person or team using, at least in part, the artifacts provided by the original authors. In our evaluation, we focused on this category by executing the complete source code from each repository to verify reproducibility. We did not consider the \textit{Results Validated - Replicated} badge, as replication requires an independent reimplementation using different methods or conditions.

\end{description}

\rowcolors{2}{white}{gray!25}
\begin{table}[htbp]  
\centering
\caption{Curated dataset}
\label{tab:curatedDataset}
\rowcolors{2}{white}{gray!25}
\begin{tabular}{m{1cm}>{\centering}m{1cm}>{\centering}m{3cm}>{\centering\arraybackslash}m{2cm}}
\toprule
\rowcolor{gray!50}
    ID/ Repo. Link & Pub. & PL/Database  & Reproduced  \\
\midrule
     \href{https://github.com/STAR-Laboratory/Accelerating-RecSys-Training}{\underline{E1}} & \cite{Adnan2022} & Unix Shell, Python& \includegraphics[width=0.4cm]{images/chap4/E-R.png} \includegraphics[width=0.4cm]{images/chap4/reproduced.png}  \\
     \href{https://github.com/jiyangbai/TaGSim}{\underline{E2}} & \cite{Bai2022} & Python& \includegraphics[width=0.4cm]{images/chap4/E-R.png} \includegraphics[width=0.4cm]{images/chap4/reproduced.png} \\
     \href{https://github.com/idea-iitd/RQuBE}{\underline{E3}} & \cite{Chauhan2022} & C++& \includegraphics[width=0.4cm]{images/chap4/E-R.png} \includegraphics[width=0.4cm]{images/chap4/reproduced.png}\\
     \href{https://github.com/jt-zhang/CardinalityEstimationTestbed}{\underline{E4}} & \cite{Sun2022}  & Python /PostgreSQL  & \includegraphics[width=0.4cm]{images/chap4/E-R.png} \includegraphics[width=0.4cm]{images/chap4/reproduced.png}\\
     \href{https://github.com/AlexanderTZhou/IUBFC}{\underline{E5}} & \cite{Zhou2022} & C++ & \includegraphics[width=0.4cm]{images/chap4/E-R.png} \includegraphics[width=0.4cm]{images/chap4/reproduced.png} \\
     \href{https://github.com/OpsPAI/ADSketch}{\underline{E6}} & \cite{Chen2022}  & Python& \includegraphics[width=0.4cm]{images/chap4/E-R.png} \includegraphics[width=0.4cm]{images/chap4/reproduced.png}\\
     \href{https://github.com/neu-se/CONFETTI}{\underline{E7}} & \cite{Kukucka2022} & Java, Python, Unix Shell & \includegraphics[width=0.4cm]{images/chap4/E-R.png} \includegraphics[width=0.4cm]{images/chap4/reproduced.png}\\
     \href{https://github.com/ICSE2022FL/ICSE2022FLCode}{\underline{E8}} & \cite{Xie2022}  & Python   & \includegraphics[width=0.4cm]{images/chap4/E-R.png} \includegraphics[width=0.4cm]{images/chap4/reproduced.png}\\
    \href{https://zenodo.org/record/4588383#.ZEwRHnbMJD8}{\underline{E9}} & \cite{Jehn2021} & Python  & \includegraphics[width=0.4cm]{images/chap4/E-R.png}  \includegraphics[width=0.4cm]{images/chap4/Available.png} \includegraphics[width=0.4cm]{images/chap4/reproduced.png}  \\
    \href{https://zenodo.org/record/7613549#.ZEwQSnbMJD8}{\underline{E10}} & \cite{Lin2024} & Python & \includegraphics[width=0.4cm]{images/chap4/E-R.png}  \includegraphics[width=0.4cm]{images/chap4/Available.png} \includegraphics[width=0.4cm]{images/chap4/reproduced.png}\\
     \href{https://zenodo.org/record/7595510}{\underline{E11}} & \cite{Santucci2023} & Python  & \includegraphics[width=0.4cm]{images/chap4/E-R.png}  \includegraphics[width=0.4cm]{images/chap4/Available.png} \includegraphics[width=0.4cm]{images/chap4/reproduced.png}\\
     
    \href{https://chicago.github.io/food-inspections-evaluation/}{\underline{E12}} & \cite{e26}  & R  & \includegraphics[width=0.4cm]{images/chap4/E-R.png} \includegraphics[width=0.4cm]{images/chap4/reproduced.png}\\
    \href{https://bitbucket.org/TonHai/iqe/src/master/}{\underline{E13}} & \cite{e28}   & Python /SQLite  & \includegraphics[width=0.4cm]{images/chap4/E-R.png} \includegraphics[width=0.4cm]{images/chap4/reproduced.png}\\
     \href{https://zenodo.org/records/3980817}{\underline{E14}} & \cite{e29}   & Jupyter Notebook & \includegraphics[width=0.4cm]{images/chap4/E-R.png}  \includegraphics[width=0.4cm]{images/chap4/Available.png} \includegraphics[width=0.4cm]{images/chap4/reproduced.png} \\
     \href{https://zenodo.org/records/4562932}{\underline{E15}} & \cite{e30}   & R & \includegraphics[width=0.4cm]{images/chap4/E-R.png}  \includegraphics[width=0.4cm]{images/chap4/Available.png} \includegraphics[width=0.4cm]{images/chap4/reproduced.png} \\
     \href{https://zenodo.org/records/3377463}{\underline{E16}} & \cite{e32}  & R/Python/Unix Shell & \includegraphics[width=0.4cm]{images/chap4/E-R.png}  \includegraphics[width=0.4cm]{images/chap4/Available.png} \includegraphics[width=0.4cm]{images/chap4/reproduced.png} \\
     \href{https://zenodo.org/records/5656923}{\underline{E17}} & \cite{e33}  & R  & \includegraphics[width=0.4cm]{images/chap4/E-R.png}  \includegraphics[width=0.4cm]{images/chap4/Available.png} \includegraphics[width=0.4cm]{images/chap4/reproduced.png} \\
    \href{https://zenodo.org/records/8270217}
    {\underline{E18}} & \cite{e38} & Python & \includegraphics[width=0.4cm]{images/chap4/E-R.png}  \includegraphics[width=0.4cm]{images/chap4/Available.png} \includegraphics[width=0.4cm]{images/chap4/reproduced.png} \\
        \bottomrule
\end{tabular}
\end{table}

\section{Threats to Validity} \label{sec:datasetThreats}
In this section, we discuss several potential threats to the validity of the study we conducted, which involves the collection and evaluation of experiments. These threats are categorized into four main areas—construct validity, internal validity, external validity, and reliability—each of which is addressed subsequently~\cite{Cook1979, Wohlin2012}.

\paragraph{Construct Validity} A potential threat is the selection bias in assembling the dataset. While we aimed for diversity by including experiments across multiple disciplines and PLs, the chosen experiments may not fully represent all computational research scenarios. Additionally, our reliance on publicly available repositories may introduce bias. Such repositories are often curated by authors who prioritize certain features or domains. We addressed this threat by using systematic criteria to select experiments from diverse domains, ensuring the diversity of scientific fields and programming environments.

\paragraph{Internal Validity} There are concerns about the integrity of the experiments collected and potential errors in assembling the dataset. Threats include the accuracy of the computational environment, completeness of dependency information, and correctness of execution instructions provided with the experiments.
Each experiment in the curated dataset was thoroughly reviewed to ensure accurate and complete information. However, inconsistencies or omissions in the original repositories might persist. To mitigate this, we resolved all the issues, such as incomplete dependency descriptions and incomplete PL descriptions.

\paragraph{External Validity}
Although the curated dataset spans multiple disciplines and computational setups, the experiments primarily focus on academic and open-source experiments. This focus may limit the applicability of the findings to proprietary or industry-driven research environments, which may involve different computational workflows.
We acknowledge this limitation and propose this curated dataset as a starting point for evaluating reproducibility tools in academic settings. Future work could extend the dataset to include more industry-oriented experiments. 

\paragraph{Reliability} Variations in how experiments are executed across different computational environments (\eg, different Operating System (OS), hardware configurations, or software versions) can pose challenges to reproducibility. These differences may impact the results when using the dataset to assess the performance and robustness of reproducibility tools.
To mitigate this threat, we standardized the preparation and documentation process for the experiments on the curated dataset. Local tools were executed within a single, consistent computational environment. For experiments involving online tools, we established new environments for each test to closely simulate the intended computational environments, we ensured that the setup was thoroughly documented, including details on software dependencies, configurations, and execution procedures. Where feasible, we created reproducibility packages using multiple tools to facilitate future reproducibility efforts. These packages, along with their information, are publicly available in the Zenodo repository to serve as a consistent reference~\cite{costa_2025_packahe_experiments_anonymous}.
%

\section{Discussion} \label{sec:datasetDiscussion}
The development of this curated dataset highlights the critical role of standardized benchmarks in evaluating reproducibility tools. Our findings emphasize the challenges that computational researchers face, particularly in ensuring that experiments are well-documented, executable across diverse environments, and supported by existing reproducibility tools.

One of the most striking observations is the significant gap between the number of experiments collected and those successfully reproduced. Of the 38 experiments gathered, only 18 (47\%) could be executed using at least one reproducibility tool. The primary barriers included missing data or files (16\%) and software version incompatibilities (10\%). This reinforces the necessity for standardized documentation practices, including explicit dependency lists and detailed execution instructions, to support reproducibility efforts.

Furthermore, the dataset reveals disparities in the effectiveness of existing reproducibility tools. While some tools excel at handling Python-based experiments, they struggle with more complex setups, such as experiments requiring specialized databases, multi-language implementations, or older software dependencies. No single tool was capable of executing all the experiments, indicating that current solutions still have substantial limitations. This finding underscores the need for more adaptable and comprehensive tools capable of supporting the diverse requirements of computational research.

A key takeaway from this study is the importance of thoroughly documenting the computational environment and execution details. Experiments that provided detailed computational information—such as specific software environments, precise dependency versions, and clear execution steps—were significantly more likely to be successfully reproduced. This suggests that future efforts in reproducibility should emphasize not only making data and code available but also ensuring that experiments are accompanied by the computational environment and execution details.

Looking ahead, this dataset serves as a foundation for further advancements in reproducibility research. Future iterations should aim to include a broader spectrum of experiments, incorporating industry-driven computational research alongside academic experiments. Additionally, expanding the dataset to cover more computational environments—such as cloud-based platforms and high-performance computing clusters—could help further assess the adaptability of reproducibility tools.

By addressing these challenges, this curated dataset not only provides a robust benchmark for evaluating reproducibility tools but also contributes to the broader effort of fostering transparency, reliability, and collaboration in computational research.

\section{Conclusion} \label{sec:datasetConclusion}
This work introduces a meticulously curated dataset designed to assess reproducibility tools in computational research. By analyzing a variety of computational experiments, we have highlighted significant gaps in documentation practices, underscoring the urgent need for improvements to ensure reliability and transparency in scientific research.

Our findings reveal that, despite advancements in reproducibility practices, substantial challenges remain, particularly regarding dependency management and environmental compatibility. With only a 47\% success rate in experiment reproducibility, the importance of advancing toward more robust and adaptable solutions is clear.

Furthermore, it is imperative that the scientific community commit to adopting more rigorous and transparent documentation practices. Detailed and accessible documentation is essential to enable researchers to effectively verify and reproduce each other's work, thereby improving integrity and accountability in research.

This work highlights the importance of reproducibility in computational research and serves as a wake-up call for researchers to collaborate in the continuous improvement of scientific practices. This dataset serves as a practical resource for evaluating and improving reproducibility tools. Based on the knowledge derived from this work, future efforts can significantly improve the reliability, transparency, and accessibility of computational research.


\balance
\bibliographystyle{ACM-Reference-Format}
\bibliography{bibliography}

\end{document}